\documentclass[conference,a4paper]{IEEEtran}
\usepackage{graphicx,xcolor,tikz}
\usepackage{comment}
\usepackage{subfig}
\usepackage{float}
\usepackage{multicol}
\usepackage{cite}
\usepackage{makecell}
\usepackage{siunitx}
\usepackage{multirow}
\usepackage{mathtools}
\usepackage{amsmath,amssymb, amsfonts}
\usepackage{tabulary}

\usepackage{cite}

\usepackage{mathtools}

\def\nb0{{\mathbf{0}}}
\def\nb1{{\mathbf{1}}}

\usepackage{tikz,epstopdf}

\usetikzlibrary{arrows}
\usepackage{caption}
\usepackage{subfig}

\usepackage{lipsum,graphicx}
\usepackage[para]{threeparttable}
\usepackage{epsfig,graphics,graphicx,subfig,amssymb,amstext,amsmath,algorithm, algorithmic,wrapfig,multirow}
\usepackage{adjustbox}
\usepackage{siunitx}
\usepackage{pgfplots}
\usepackage{xcolor}
\usepackage{url}

\usepackage{breakurl}
\usepackage[breaklinks]{hyperref}
\usetikzlibrary{chains,arrows,calc ,positioning}
\usepackage{graphicx}
\usepackage{colortbl, xcolor}
\usepackage{enumitem}
\setitemize{noitemsep,topsep=0pt,parsep=0pt,partopsep=0pt}
\definecolor{maroon}{cmyk}{0,0.87,0.68,0.32}

\pgfplotsset{compat=1.16}
\definecolor{bittersweet}{rgb}{1.0, 0.44, 0.37}
\definecolor{glaucous}{rgb}{0.38, 0.51, 0.71}
\definecolor{gainsboro}{rgb}{0.86, 0.86, 0.86}
\definecolor{babyblueeyes}{rgb}{0.63, 0.79, 0.95}
\definecolor{silver}{rgb}{0.75, 0.75, 0.75}
\definecolor{neoncarrot}{rgb}{1.0, 0.64, 0.26}

\def\INR    {{\mathsf{INR}}}
\newcommand{\wh}[1]{\widehat{{#1}}}
\newcommand{\bsym}[1]{\boldsymbol{{#1}}}
\newcommand{\subsf}{\sf \scriptscriptstyle}
\newcommand{\Herm}{{\subsf H}}

\newcommand*{\rom}[1]{\expandafter\@slowromancap\romannumeral #1@}

\usepackage[normalem]{ulem}

\makeatletter

\providecommand{\tabularnewline}{\\}

 \let\oldforeign@language\foreign@language
 \DeclareRobustCommand{\foreign@language}[1]{%
   \lowercase{\oldforeign@language{#1}}}

\usepackage[font=small]{caption}

\usepackage{dblfloatfix}

\makeatother

\IEEEoverridecommandlockouts

\begin{document}

\bstctlcite{IEEEexample:BSTcontrol}

%
\title{Terrestrial-Satellite Spectrum 
Sharing in the\\Upper Mid-Band with
Interference Nulling}
%

\author{
%
\IEEEauthorblockN{Seongjoon Kang$^{\dagger}$  \quad Giovanni Geraci$^{\sharp}$ 
 \quad Marco Mezzavilla$^{\dagger}$  \quad Sundeep Rangan$^{\dagger}$ 
\vspace{0.1cm}} 
\IEEEauthorblockA{$^{\dagger}$NYU Tandon School of Engineering, Brooklyn, NY, USA}
\IEEEauthorblockA{$^{\sharp}$Telef\'{o}nica Research and Universitat Pompeu Fabra, Barcelona, Spain}

\thanks{S. Kang, M. Mezzavilla, and S. Rangan were supported by NSF grants 1952180, 2133662, 2236097, 2148293, 1925079, the NTIA,  and the industrial affiliates of NYU WIRELESS. G. Geraci was supported by the Spanish Research Agency grants PID2021-123999OB-I00 and CEX2021-001195-M, by the UPF-Fractus Chair, and by the Spanish Ministry of Economic Affairs and Digital Transformation and the European Union NextGenerationEU through the UNICO 5G I+D SORUS project.}
}

\maketitle

\begin{abstract}
The growing demand for broader bandwidth in cellular networks has turned the upper mid-band (7--24 GHz) into a focal point for expansion. 
However, the integration of terrestrial cellular and incumbent satellite services, particularly in the 12\,GHz band, poses significant interference challenges. This paper investigates the interference dynamics in terrestrial-satellite coexistence scenarios and introduces a novel beamforming approach that leverages available ephemeris data for dynamic interference mitigation. By establishing spatial radiation nulls directed towards visible satellites, our technique ensures the protection of satellite uplink communications without markedly compromising terrestrial downlink quality. Through a practical case study, we demonstrate that our approach maintains the satellite uplink signal-to-noise ratio (SNR) degradation under 0.1\,dB and incurs only a negligible SNR penalty for the terrestrial downlink. 
Our findings offer a promising pathway for efficient spectrum sharing in the upper mid-band, fostering a concurrent enhancement in both terrestrial and satellite network capacity.


\end{abstract}


\IEEEpeerreviewmaketitle

\section{Introduction}

The upper mid-band from 7--24 \si{GHz} has attracted considerable interest
to expand cellular services \cite{kang2023cellular}.
These frequencies offer much more
abundant spectrum than that available in 
the congested sub-7\,\si{GHz} 
bands with more favorable
propagation and coverage than the
mmWave frequencies.  With this balance
of coverage and spectrum, the upper mid-band has become the focus of significant work in industry 
\cite{smee2022ten,samsung20226gspectrum}, the Federal Communications Commission (FCC) \cite{fcc2023preliminary,fcc20236Gworkinggroup}, and the 3rd Generation Partnership Project (3GPP) \cite{3GPP38921}.

A key challenge for expanding
cellular service in the upper mid-band
is to coexist with incumbent services, particularly
commercial communication satellites. 
Mega-constellations of Low Earth Orbit (LEO) satellites can enable new services and provide vital high-capacity coverage
in rural areas where fixed infrastructure is costly \cite{LinCioCha2021,GerLopBen2023,BenGerLop2022,LeySorRop2020}.  
Starlink, currently the largest
service provider, has deployed approximately
$4700$ LEO satellites, as of September 2023, and plans to expand to over 25000 \cite{PacPorCra21}. 
The upper mid-band is a key part
of Starlink's service expansion as shown, for instance, by a recent FCC filing \cite{fcc-starlink-2022}.

Operating cellular networks in the upper mid-band may disrupt incumbent satellite service links \cite{yastrebova2020theoretical, deslandes2010analysis}.
In particular, as cellular base stations (BSs) employ large antenna arrays and high transmit power, the sidelobes generated by their beamformed transmissions toward the user equipment (UEs) can result in strong interference to the satellite uplink. 
While early attempts were made at mitigating cellular-to-satellite interference through null-steering \cite{jo2011transmit,lim2007interference}, these approaches focus on one-dimensional beam spaces and are not directly applicable to the rectangular array configurations of current BSs. Morevoer, they may incur a significant beamforming gain loss when the nulling directions are close to the main serving direction. 

\begin{figure}[t]
\centering
{
\includegraphics[width=0.85\columnwidth]{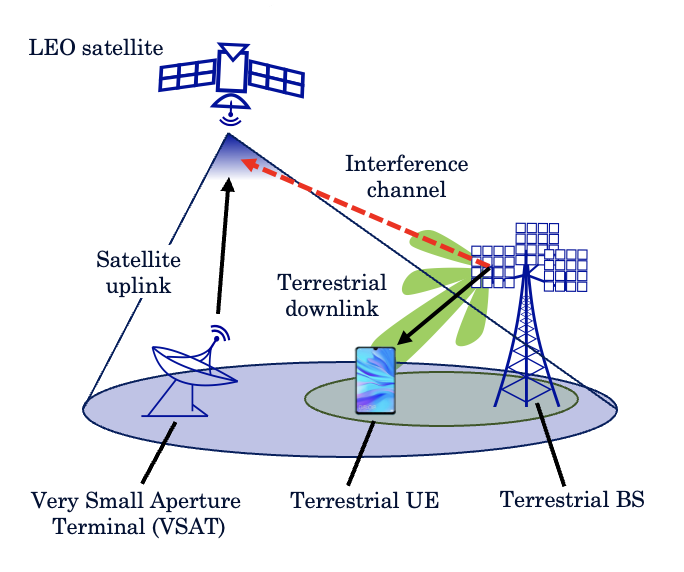}
}
 \caption{Illustration of the interference caused by the cellular downlink to the satellite uplink through BS antenna sidelobes.} 
 \label{fig:sidelob_itf}
\end{figure}

In this work, we consider practical cellular deployments in the upper mid-band with realistic antenna configurations and present a novel approach to mitigate cellular-to-satellite interference. In particular, we focus on the \SI{12}{GHz} band, given the growing interest it has attracted thanks to its abundant available spectrum and benign propagation \cite{hassan2023spectrum}.%
\footnote{Note that the FCC released a proposed rulemaking notice (NPRM) to discuss the spectrum sharing of the \SI{12}{GHz} band with 5G systems \cite{fcc2023preliminary}.}
Our proposed approach leverages LEO satellite tracking information using \emph{ephemeris} data, and it aims to mitigate interference generated by the downlink of a cellular network to the uplink of the LEO satellite as shown in Fig.~\ref{fig:sidelob_itf}, since this represents the worst case. 
Our main contributions are as follows:
\begin{itemize}    
    \item \emph{Cellular-to-satellite interference analysis:} We analytically illustrate the relationship between the satellite elevation angle and the corresponding cellular-to-satellite pathloss, identifying ranges susceptible to interference.
    \item \emph{Proposed interference nulling}: 
    We propose an interference nulling method in which terrestrial beamforming is carried out while creating spatial radiation nulls onto certain directions based on line-of-sight (LOS) channels computed from practically available ephemeris data.
    \item \emph{Case study}: We track Starlink satellites from a sample rural area near Boulder, Colorado, and evaluate terrestrial-satellite spectrum sharing in the upper mid-band. We demonstrate that the proposed interference nulling method can consistently keep the degradation in the satellite uplink signal-to-noise ratio (SNR) below \SI{1}{dB} while only penalizing the terrestrial downlink SNR by \SI{0.1}{dB} in median.
\end{itemize}

\section{Single Victim Satellite in a LOS Channel}
\label{sec:los_itf_channel}

To understand the nature of the interference
from the terrestrial downlink to the satellite
uplink, we first consider the case of 
a single victim satellite in a LOS scenario
as illustrated in Fig.~\ref{fig:sidelob_itf}.
We will analyze multiple satellites and non-LOS
channels later.
The interference in this scenario is 
influenced by two primary factors: the BS antenna element gain and the satellite elevation angle. On the one hand, when satellites are positioned at higher elevation angles, they are typically situated more directly above the BSs. Consequently, the distance between satellites and BSs decreases and so does the path loss, potentially intensifying interference to the detriment of the satellites. On the other hand, since BS antennas are generally slightly down-tilted, their antenna gains diminish at higher elevation angles, thereby mitigating interference. As a result, the level of interference experienced by any given satellite can vary significantly across different elevation angles and this variability necessitates a suitable approach to managing interference.
 
\begin{figure}[t!]
\centering
{
\includegraphics[width=0.95\columnwidth]{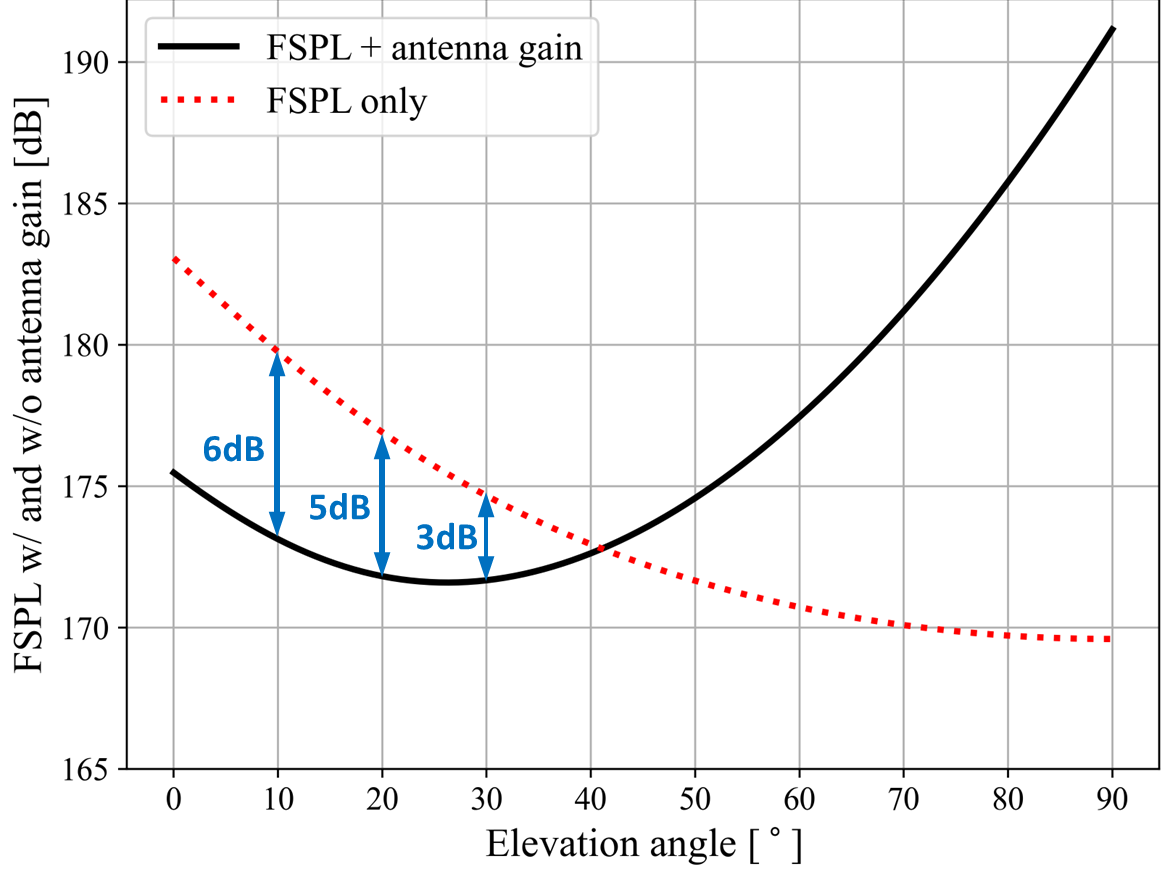}
}
 \caption{Change in the total BS-to-satellite propagation loss caused by accounting for the antenna element gain as a function of the LEO satellite elevation angle $\theta$ for an azimuth angle fixed to $\phi=0^\circ$.} 
 \label{fig:pl_cdf}
\end{figure}

The separation between the target satellite and the BS, termed the \emph{slant distance}, is given as follows
\begin{align} \label{eq:slant_dist}
    d(\theta) = \sqrt{R_E^2\sin^2(\theta) + h^2 + 2hR_E},
    -R_E\sin(\theta)
\end{align}
where $R_E$ denotes the Earth's radius, $h$ is the height of BS, and $\theta$ is the elevation angle. It can be observed that the slant distance increases as the elevation angle $\theta$ in the global coordinate system decreases. Given $d(\theta)$, the free space path loss (FSPL) at a carrier frequency $f$ can be expressed as 
\begin{align} \label{eq:fspl}
    {\rm FSPL}(\theta) = 20\log_{10}(d(\theta))+20\log_{10}(f)-147.55.
\end{align}
The red dotted curve in Fig.~\ref{fig:pl_cdf} depicts the variation in path loss as a function of the satellite elevation angle $\theta$, with the azimuth angle held constant at $\phi=0^\circ$.

To characterize the sidelobes of BS antenna arrays, we consider solely the antenna element gain, as the gains due to beamforming vary depending on the specific transmission directions. We adopt the 3GPP antenna field pattern defined in \cite{3GPP37840}, described by
\begin{align}\label{eq:elem_gain}
    G_E(\theta, \phi)&=G_{\max }\!-\!\min \left\{\!-\!\left[G_{E, V}(\theta)\!+\!G_{E, H}(\phi)\right]\!, A_{\mathrm{m}}\right\} \\
    G_{E, V}(\theta)&=-\min \left\{12\left[\left(\theta-90^\circ\right)/\theta_{3\mathrm{dB}}\right]^2, SLA_{\mathrm{V}}\right\}\\
    G_{E, H}(\phi)&=-\min \left\{12\left[\phi/\phi_{3\mathrm{dB}}\right]^2, A_{\mathrm{m}}\right\},
\end{align}
and where $G_{\max }=$ \SI{8}{dB} is the maximum gain, $SLA_{\mathrm{V}} = $ \SI{30}{dB} denotes the side-lobe level limit, $\phi_{\mathrm{3dB}} = 65^\circ$ and $\theta_{\mathrm{3dB}}  = 65^\circ $  are the horizontal and vertical \SI{3}{dB} beamwidths, respectively, and $A_{\mathrm{m}} = $ \SI{30}{dB} represents the front-to-back ratio. Note that $\theta$ and $\phi$ are the elevation and azimuth angles of the BS expressed in the local coordinate system. 

For the purpose of this analysis, we focus on the impact of the BS antenna sidelobes on the LOS channels between the BS and the LEO satellite. We also assume that the element gain of the satellite remains nearly at its maximum value, representing a worst-case scenario. The solid black curve in Fig.~\ref{fig:pl_cdf} shows the variation in propagation loss vs. satellite elevation angle when accounting also for the antenna element gain in (\ref{eq:elem_gain}). We observe that, for elevation angles below $40^\circ$, the BS antenna sidelobes further reduce the propagation loss.
Given the higher likelihood of encountering lower elevation angles \cite{li2002analytical}, Fig.~\ref{fig:pl_cdf} suggests that LOS channels between BSs and satellites may pose a significant interference risk to uplink satellite communications.




\section{Proposed Interference Nulling Technique}
\label{sed:tr_based_null}

We now consider the case where
multiple victim satellites suffer interference by potential
\emph{multi-path} channels from a BS.  
As before, we focus on downlink transmission (BS $\rightarrow$ UE). 
The BS has $N_t$ transmit antennas and
the UE has $N_r$ receive antennas,
and we let $\bsym{H}_{\rm ter} \in \mathbb{C}^{N_r \times N_t}$ denote the multiple-input multiple-output (MIMO) channel matrix
from the BS to the UE, which can be frequency selective due to multi-path fading. 
We are concerned with the sidelobe interference to $N_{\rm sat}$ satellites,
indexed $i=1,\ldots,N_{\rm sat}$.
Each potential victim satellite is assumed to have some beam focused on the area covering the BS
which is sensitive to the transmission from
the BS.  We let $\bsym{h}_i \in \mathbb{C}^{N_t \times 1}$ denote the multi-path channel vector from the terrestrial
BS to the $i$-th victim satellite, which can be frequency dependent.   


\subsubsection*{Multi-path interference nulling}
The terrestrial BS and terrestrial UE 
must select, respectively, transmit and receive beamforming vectors 
$\bsym{w}_t$ and $\bsym{w}_r$.
The first algorithm we consider is \emph{multi-path interference nulling} where the beamforming vectors are selected by solving the following problem: 
\begin{equation}
\begin{aligned} \label{eq:bf_eq}
    {\bsym{w}}_r^{\rm opt}, {\bsym{w}}_t^{\rm opt} &= \arg\! \max_{\bsym{w}_r, \bsym{w}_t} \left[ 
        |\bsym{w}_r^\Herm\Tilde{\bsym{H}}_{\rm ter} \bsym{w}_t|^2 - \lambda \sum_{i=1}^{N_{\rm sat}} \lVert \Tilde{\bsym{h}}_{i}^\Herm\bsym{w}_t\lVert^2\right], \\
       &\textrm{s.t.} \quad \lVert \bsym{w}_t \rVert^2 = 1 \quad \text{and} \quad ~\lVert \bsym{w}_r \rVert^2 = 1,
\end{aligned}
\end{equation}
where $\Tilde{\bsym{H}}_{\text{ter}}$ and $\Tilde{\bsym{h}}_i$ are the normalized versions of MIMO multi-path channels $\bsym{H}_{\text{ter}}$ and $\bsym{h}_i$, respectively, and 
$\lambda \geq 0$ denotes a regularization parameter trading off beamforming gain at the terrestrial UE for nulling accuracy at the satellites.
Normalizing channel matrices ensures $\lVert\Tilde{\bsym{H}}_{\rm ter}\rVert_{F}^2 = N_t N_r$ and $\lVert\Tilde{\bsym{h}}_{i}\rVert^2 = N_t$ \cite{heath2018foundations}
and circumvents the pronounced path loss disparity between terrestrial and satellite channels, thus restricting the range of values of interest for $\lambda$. 
We approximately solve the optimization problem (\ref{eq:bf_eq})
by selecting $\bsym{w}_r^{\rm opt}$
as the maximum left singular vector of $\Tilde{\bsym{H}}_{\rm ter}$.  Then,  $\bsym{w}_t^{\rm opt}$ is given by  maximum eigenvector of 
\begin{equation}
\begin{aligned} 
\Tilde{\bsym{H}}^\Herm_{\rm ter}\bsym{w}_r\bsym{w}_r^\Herm\Tilde{\bsym{H}}_{\rm ter} - \lambda\sum_{i=1}^{N_{sat}} \Tilde{\bsym{h}}_i\Tilde{\bsym{h}_i}^\Herm
\end{aligned}
\end{equation}
Due to a multi-path environment,  the solutions of the problem (\ref{eq:bf_eq}), i.e., ${\bsym{w}}_r^{\rm opt}$ and ${\bsym{w}}_t^{\rm opt}$, depend on frequency.

\subsubsection*{Proposed LOS interference nulling}
A practical challenge in implementing  
multi-path interference nulling is that the BS must keep track of the entire 
wideband channel $\bsym{h}_i$ to each satellite. 
To avoid the complexity of real-time multi-path channel estimation between each BS and all satellites, we propose a simple \emph{LOS nulling}
method.
In this method, we replace $\Tilde{\bsym{h}}_i$
with $\bsym{e}(\theta_i,\phi_i)$ where $(\theta_i,\phi_i)$ are the azimuth
and elevation angle to the $i$-th victim
satellite and $\bsym{e}(\theta_i,\phi_i)$
is the spatial signature of the BS antenna array in that direction.
This formulation is motivated by the power predominance of LOS paths with respect to NLOS paths. 
Furthermore, all ephemeris data from commercial satellites are publicly available. Therefore, the channel vector $\bsym{e}(\theta_i,\phi_i)$ can be obtained at each BS using these public data for real-time satellite tracking, along with the respective elevation and azimuth angles. 

\subsubsection*{Example of beamforming pattern}
\begin{figure}[t!]
\centering
\includegraphics[width =0.95\columnwidth]{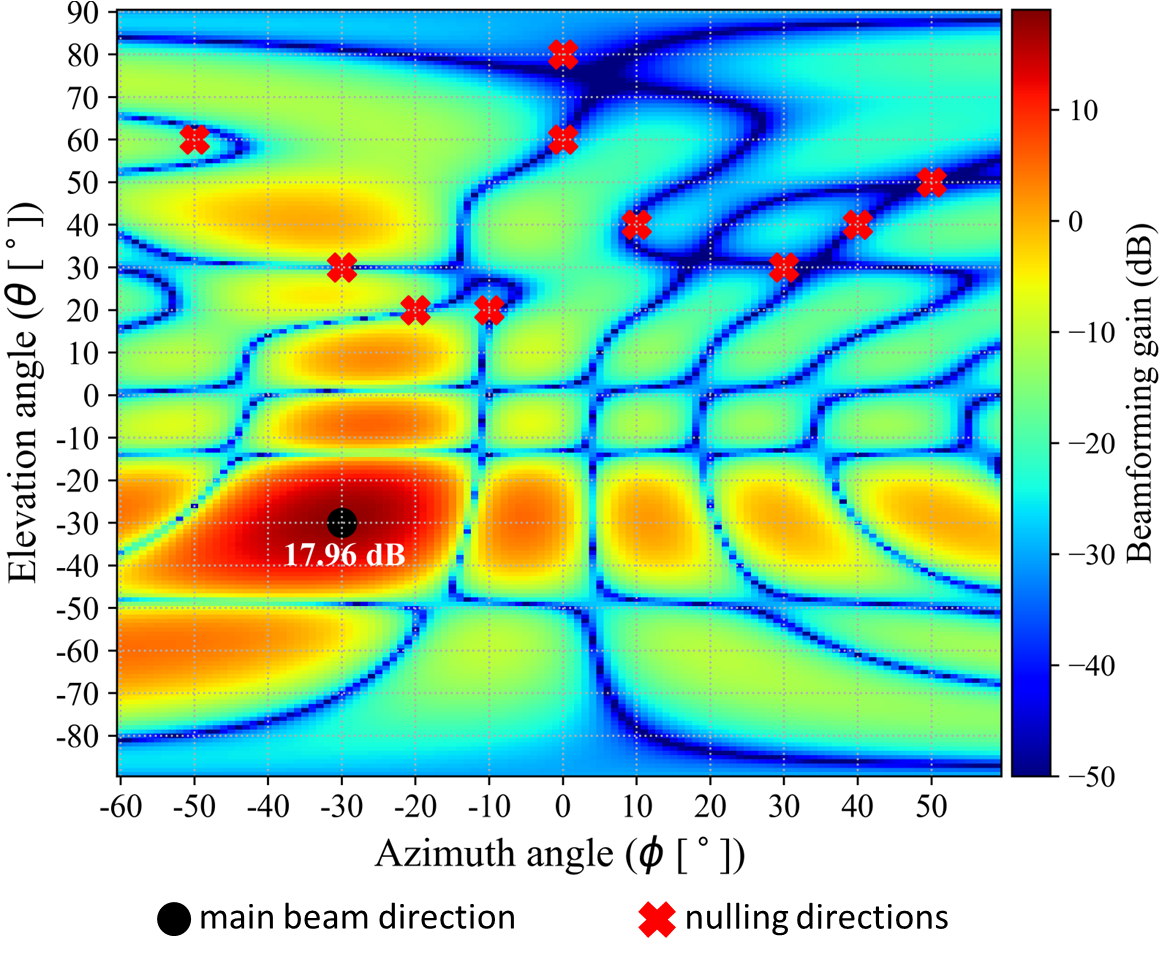}
\caption{Beamforming gains in beam space for an $8 \times 8$ BS URA.}
\label{fig:bf_gain}
\end{figure}

To illustrate the LOS beamforming method,
Fig.~\ref{fig:bf_gain} shows an example of the transmit beamforming gains 
$|{\bsym{e}^\Herm(\theta, \phi)} \bsym{w}_t|^2$ across the angular space of a BS equipped with an $8 \times 8$ uniform rectangular array (URA) 
with a vector $\bsym{w}_t$ calculated from
the proposed LOS interference nulling method. 
In this case, there are $N_{\rm sat}=10$
satellites with locations shown in the red
cross marks.  We set $\lambda = 10$ and we can see that the beamforming pattern places strong nulls at all satellite locations.  The desired destination towards the terrestrial UE 
is shown in the black circle.  In addition, the beamforming gain towards the desired UE computed with interference nulling is \SI{17.96}{dB}.  The theoretical maximum beamforming gain is $10\log_{10}(64)=$\SI{18.06}{dB}, so the loss, at least in this case,
is negligible.  

\subsubsection*{Effects of Motion}
A potential issue with either full multi-path
or LOS interference nulling is that the interference channel can vary over the transmission time of a BS.
To assess this possibility, recall that, in 5G, the transmission time interval (TTI) is as large as $T=$\,\SI{1}{ms}.
%
%
Now, a LEO satellites orbiting at an altitude of $h=$\,\SI{600}{km} would travel at a velocity of $v=\,$\SI{7.56}{km/s}. Hence, the maximum change in the the angle is at most, $\Delta \theta = \tan^{-1}\left( {vT}/{h}\right) \approx 7.2\times 10^{-4}$ degrees. 
Hence, the angular change is minimal, and LOS interference channel $\bsym{e}(\theta_i,\phi_i)$ can considered to be approximately constant over practical TTIs.

\section{Case Study and Performance Evaluation}

We now present a case study to evaluate the performance of the proposed interference nulling. The main assumptions are detailed in the sequel and summarized in Table~\ref{tab:sim_params}.


\subsection{System Model and Methodology}

\subsubsection*{Terrestrial channel model}

We focus on downlink transmission, since this represents the worst case in terms of interference caused at satellites. We consider a \SI{24}{km} $\times$ \SI{15}{km} rural area near Boulder, Colorado, illustrated in Fig.~\ref{fig:rural}, where BSs are deployed on a grid with \SI{1732}{m} inter-site distance (ISD). Terrestrial users (UEs) are deployed uniformly at random throughout the region and associate to the BS providing the largest signal strength. We consider a scenario with $20\%$ network load, resulting in 21 BSs (out of the total 104) transmitting simultaneously, each with an $8\times8$ URA. For this setup, following the 3GPP specifications in \cite{3GPP38901}, we simulate the propagation channel using the commercial ray-tracing software Wireless InSite \cite{remcom}.

\subsubsection*{Satellite channel model}

\begin{figure}[t!]
\centering
    {
    \includegraphics[width=0.95\columnwidth]{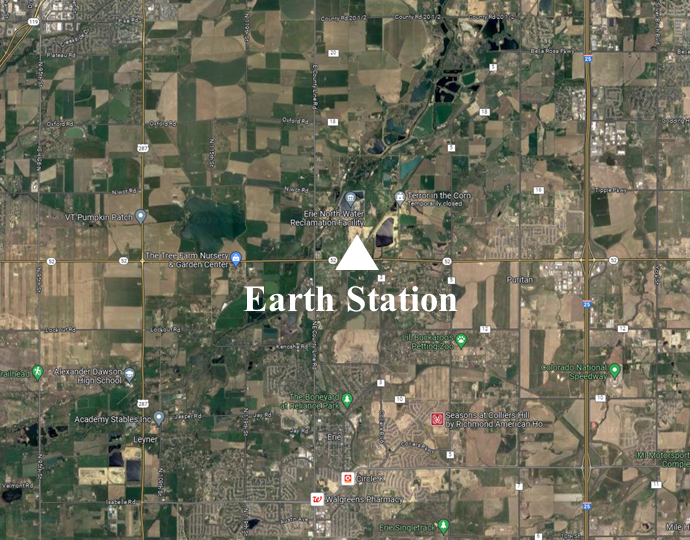}
    }
 \caption{Rural area near Boulder, CO, used for our case study.} 
 \label{fig:rural}
\end{figure}

We assume that LEO satellites in a mega-constellation are employed to provide service to satellite UEs within the same rural area, sharing the same spectrum as the terrestrial deployment in the upper mid-band. To capture the real-time positions of the satellites, we use ephemeris data in the two-line element set (TLE) format \cite{3GPP38821}, available from public databases \cite{celesTrack}, and process it through the Python library \emph{pycraf} \cite{winkel2018pycraf}. 
As shown in Fig.~\ref{fig:rural}, we consider a single Earth Station within the given rural area where satellites are tracked at every minute. 
We consider a fixed number of $\rm N_{\rm sat}$ satellites chosen randomly at each interval from the pool of visible satellites. 
For those, we obtain the multi-path channel components via ray-tracing and add the relevant attenuation effects from atmospheric conditions, rain, clouds, and scintillation using the Python library \textit{ITU-Rpy} \cite{iturpy-2017}.

\subsubsection*{Real-time channel extraction}

\begin{figure}[t!]
\centering
\includegraphics[width =0.95\columnwidth]{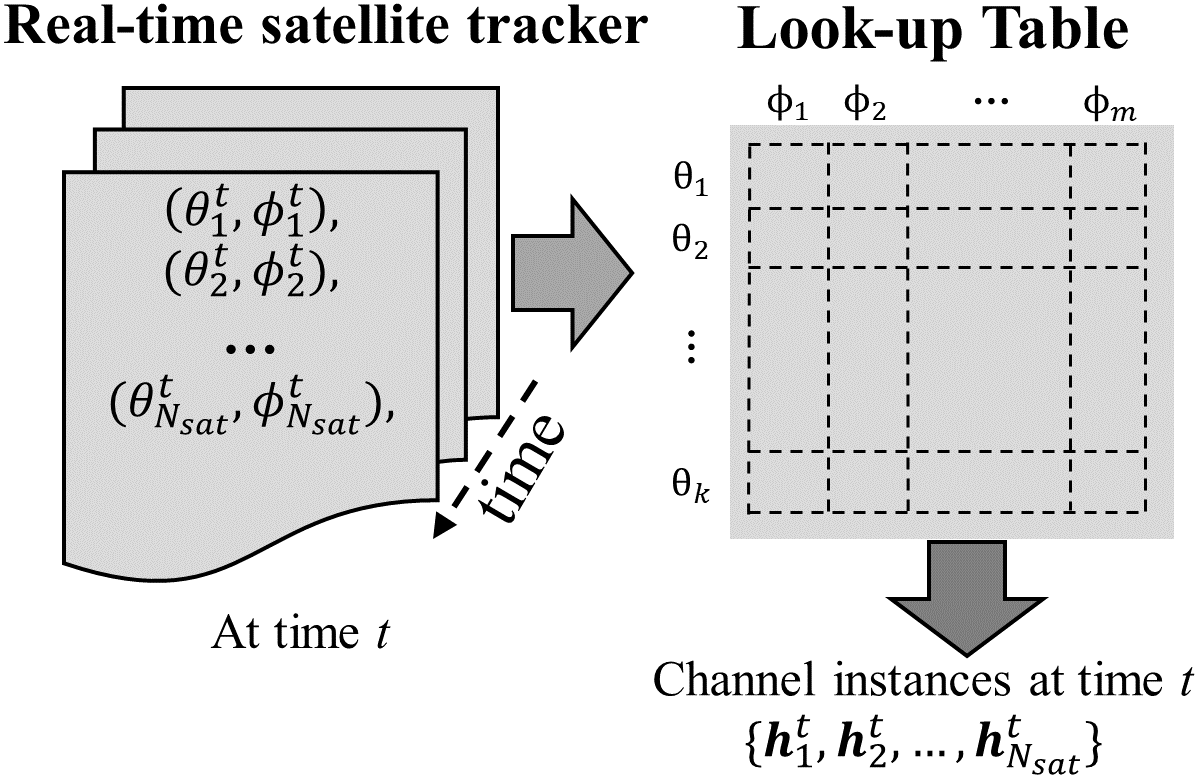}
\caption{Computation of satellite multi-path channels: (i) satellite coordinates $(\theta^{t}, \phi^{t})$ at time $t$ are tracked via ephemeris data; (ii) ray-tracing is employed to pre-compute the multi-path channels for each set of coordinates in the look-up table; and (iii) over time the multi-path channel instances are extracted from the look-up table.}
\label{fig:sat_chan}
\end{figure}

While ray-tracing simulations allow us to obtain all propagation channels, the BS-to-satellite channels vary dynamically over time due to satellite motion, and it is impractical to perform real-time ray-tracing. To overcome this issue, we pre-calculate these channels for a set of fixed positions categorized by satellite elevation and azimuth angles, creating a look-up table as shown in Fig.~\ref{fig:sat_chan}. Specifically, we include discrete points spaced by $60^\circ$ in azimuth and $10^\circ$ in elevation. This look-up table allows us to approximate real-time satellite channels by selecting entries in the table closest to the current satellite position.

\subsection{Key Performance Indicators}

\subsubsection*{Satellite interference-to-noise ratio}
The interference-to-noise ratio (INR) serves as a key metric to quantify interference in spectrum sharing scenarios. 
For a satellite $i$ and BS $j$ employing transmit beamforming vector $\bsym{w}_{j}$, the INR is calculated as follows:
\begin{equation}
    \INR_i \!=\! P_{\rm tx} + 10 \log_{10} \sum_{j=1}^{\rm N_{\rm BS}}{|\bsym{w}_j^\Herm \bsym{h}_{ij}|^2} 
    + \frac{G}{T} - L_a - 10 \log_{10}(B \cdot \kappa),
    \label{eq:inr}
\end{equation}
where $P_{\rm tx}$ denotes the BS transmit power, ${\rm N_{\rm BS}}$ is the number of concurrent BS transmissions, $\bsym{h}_{ij}$ is the multi-path interference channel between BS $j$ and satellite $i$, $G/T$ is the satellite antenna gain-to-noise-temperature ratio, $L_a$ incorporates additional losses such as atmospheric and scintillation attenuation \cite{3GPP38821}, $B$ is the bandwidth, and $\kappa$ is Boltzmann constant. Note that the INR on the satellite link can be directly mapped to an SNR degradation $\rho^{\rm S}$ as follows
\begin{align} 
\label{eq:snr_degradation_eq}
\rho^{\rm S} = 10 \log _{10}\left(1+10^{0.1 \cdot \mathrm{INR}}\right),
\end{align}
e.g., for $\rho^{\rm S} <$\,\SI{1}{dB} the INR should be kept below \SI{-6}{dB}.

\subsubsection*{Terrestrial SNR degradation}
In addition, to evaluate the loss of terrestrial SNR we define the following quantity \cite{kang2023cellular}
\begin{equation}
    \rho^{\rm T} = 10 \log_{10}\left( \frac{|{\bsym{w}}_r^{\rm opt} \bsym{H}_{\rm ter}{\bsym{w}}_t|^2}{
    |{\bsym{w}}_r^{\rm opt} \bsym{H}_{\rm ter} \wh{\bsym{w}}_t^{\lambda}|^2} \right) \nonumber
\end{equation}
where $\bsym{w}_r^{\rm opt}$ is the receive beamformer obtained from (\ref{eq:bf_eq}), whereas $\bsym{w}_t$ and $\wh{\bsym{w}}_t^{\lambda}$ are the transmit beamformers obtained from (\ref{eq:bf_eq}) for $\lambda=0$ and for $\lambda > 0$, respectively, i.e., without and with interference nulling.




\subsection{Coexistence Evaluation Without Interference Nulling}

\begin{figure}[!t]
\centering
\subfloat[][CDF of the INR on the satellite uplink without interference nulling.]
{\includegraphics[width =0.95\columnwidth]{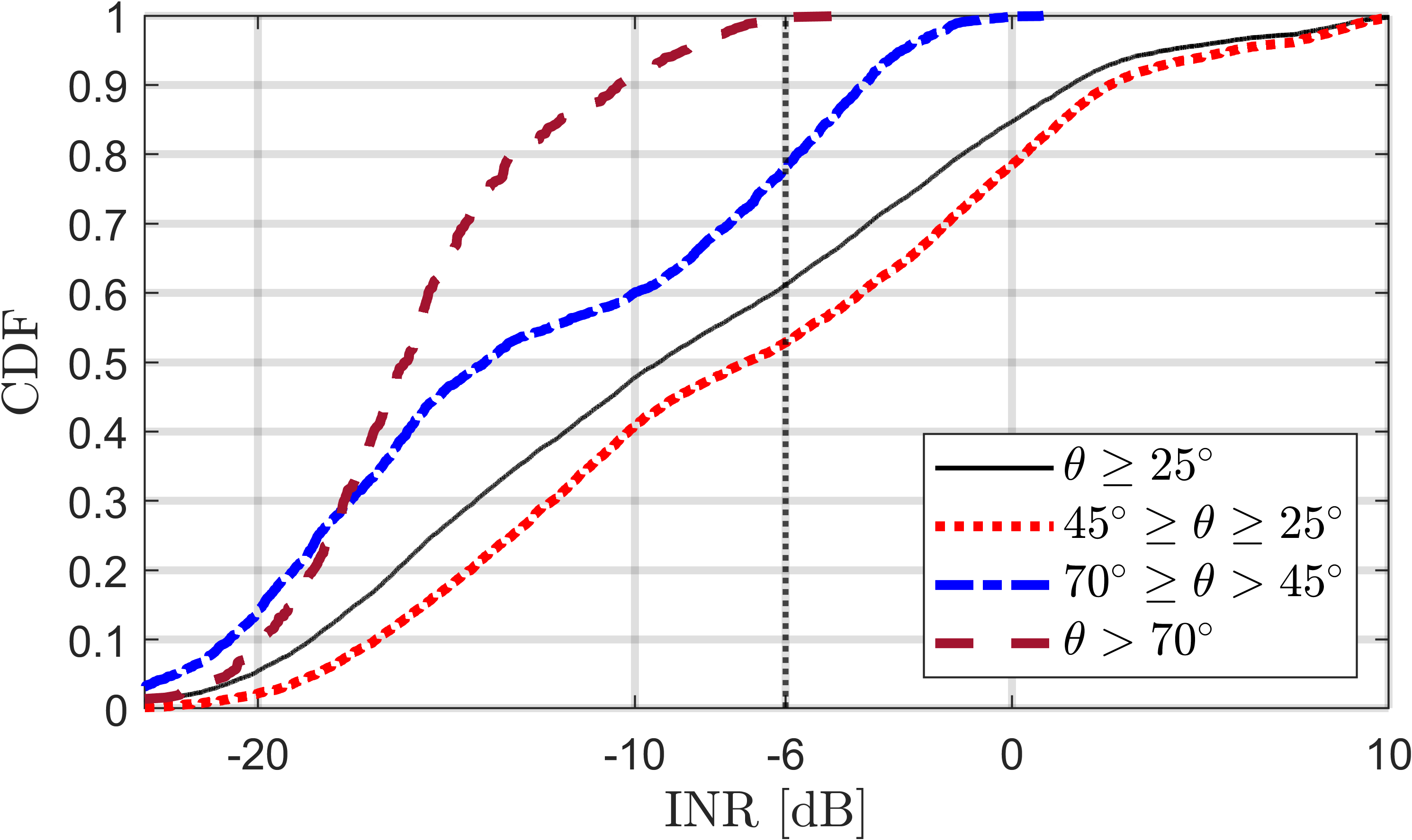} \label{fig:inr_elev_cdf}}\\
\subfloat[][PDF of the satellite elevation angles for Starlink constellation.]
{\includegraphics[width =0.95\columnwidth]{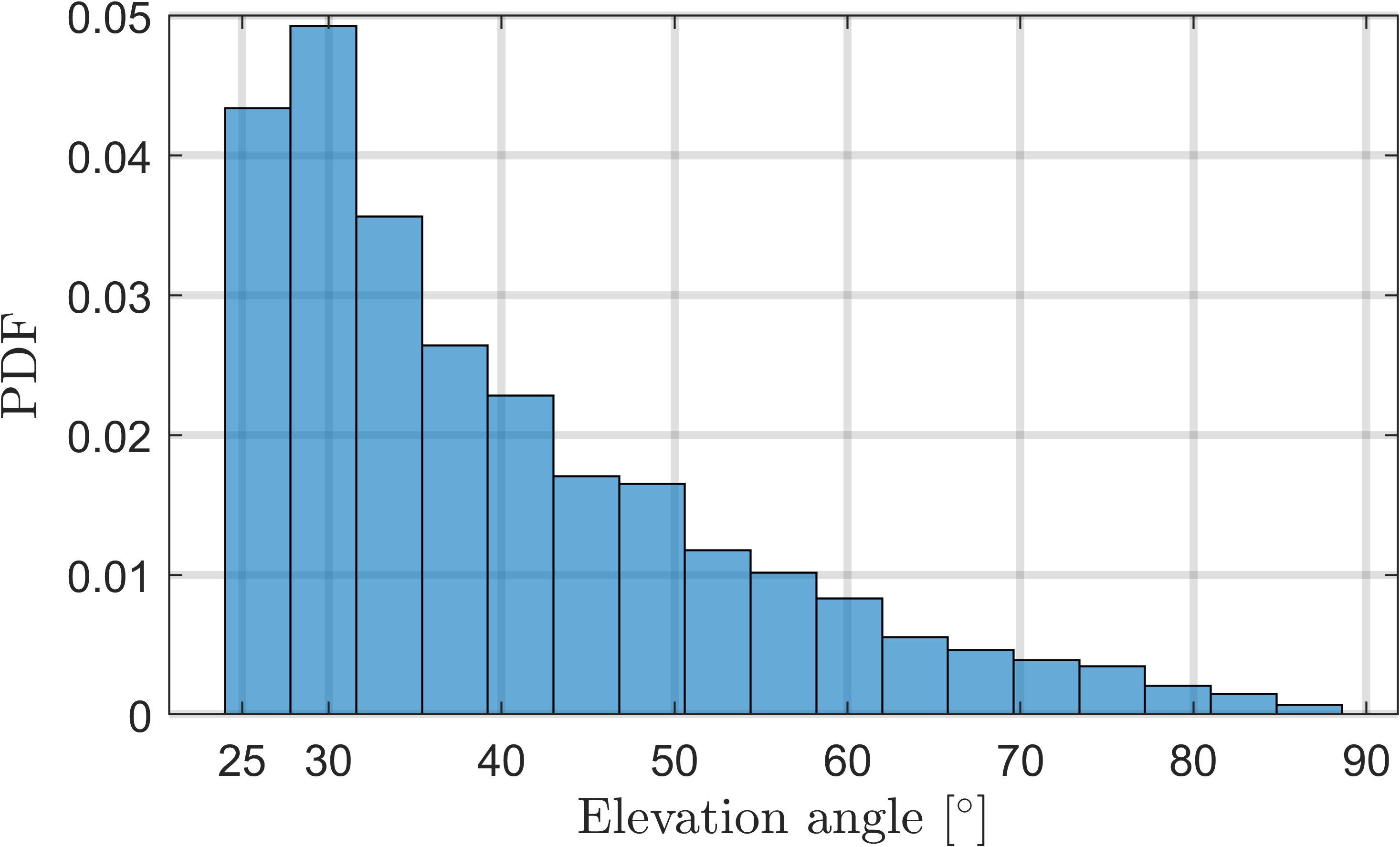} \label{fig:elev_pdf}}
\caption{(a) CDF of the INR on the satellite uplink without interference nulling vs. elevation angle range, and (b) PDF of the elevation angles.}
\label{fig:inr_pdf_nbs10}
\end{figure}

To establish a baseline, we assess the interference experienced by satellites in the absence of interference nulling, when terrestrial BSs apply singular value decomposition (SVD)-based transmit beamforming. This is achieved by resolving (\ref{eq:bf_eq}) with $\lambda=0$, thus ignoring any interference to satellites. Fig.~\ref{fig:inr_elev_cdf} illustrates the cumulative distribution function (CDF) of the INR across different satellite elevation angle ranges $\theta$: specifically, for 
$\theta \in [25^\circ, 45^\circ]$ (dotted red), $\theta \in [45^\circ, 70^\circ]$ (dash-dot blue), and $\theta \in [70^\circ, 90^\circ]$ (dashed gray). The CDF for the entire spectrum of elevation angles, i.e., $\theta \in [25^\circ, 90^\circ]$, is depicted as well (solid black).

As previously discussed in Section~\ref{sec:los_itf_channel}, lower elevation angles may lead to increased interference due to the gain from the BS antenna elements. Fig.~\ref{fig:inr_elev_cdf} specifically demonstrates that when $\theta \in [25^\circ, 45^\circ]$ and $\theta \in [45^\circ, 70^\circ]$, over $45\%$ and $20\%$ of terrestrial transmissions result in an $\text{INR}>$\SI{-6}{dB} for the satellite uplink. In contrast, a $\text{INR}>$\SI{-6}{dB} is observed less than $1\%$ of the time when $\theta > 70^\circ$. However, the likelihood of encountering satellites at lower elevation angles is greater. The probability distribution function (PDF) of the elevation angle for the satellites observed from the Starlink constellation is shown in Fig.~\ref{fig:elev_pdf}, aligning with the analytical findings in \cite{li2002analytical}. Consequently, the aggregate CDF of the INR for all elevation angles (solid black line in Fig.~\ref{fig:inr_elev_cdf}) corroborates the presence of considerable terrestrial-to-satellite interference, underscoring the importance of employing interference mitigation strategies.


\subsection{Performance Evaluation of Interference Nulling}

\begin{table}
\begin{threeparttable}
\centering
\caption{System-level simulation parameters.}
\label{tab:sim_params}
\def\arraystretch{1.2}
\begin{tabulary}{\columnwidth}{ |p{4.1cm} | p{3.85cm} | }
\hline
\textbf{Terrestrial network} &  {} \\ \hline
Geographical area & 24\,km $\times$ 15\,km near Boulder, CO \\ \hline
BS deployment & $1732$\,m ISD, 3 sectors/site \\ \hline
UE deployment & Uniform random \\ \hline
Carrier frequency and noise figure & $12$\,GHz and $7$\,dB (UE) \\ \hline
BS transmit power and network load & $33$\,dBm, 20\% of the BSs active \\ \hline
Antenna height & $35$\,m (BS) and $1.6$\,m (UE) \\ \hline
Antenna element radiation pattern & As per \cite{3GPP37840}, $12^\circ$ downtilt (BS) \\ \hline
Antenna array configuration & $8 \times 8$ (BS) and $1 \times 2$ (UE) \\ \hline\hline
\textbf{Satellite network} & \textbf{} \\ \hline
Tracked satellite constellation & Starlink, min. elevation $25^\circ$ \cite{fcc2022grantSpaceX} \\ \hline
Coordinates of the Earth station & \ang{40;04;1.12}N, \ang{105;05;15.33}W \\ \hline
Satellite bandwidth and $G/T$  & $30$\,MHz, $13$\,dB/k \\ \hline 
\end{tabulary}
\end{threeparttable}
\end{table}

We evaluate three different beamforming approaches:
\begin{itemize}
    \item  \emph{Multi-path interference nulling}: Achieved by solving (\ref{eq:bf_eq}) with the full multi-path interference channels $\Tilde{\bsym{h}}_i$. Note that in this case all the evaluation quantities are averaged in linear scale over available carrier frequencies, since the multi-path channels can be frequency selective. 
    \item \emph{LOS interference nulling}: Achieved by solving (\ref{eq:bf_eq}) employing the LOS interference channel $\Tilde{\bsym{h}}_i = \bsym{e}(\theta_i,\phi_i)$. Recall that the BSs can obtain this matrix by tracking the elevation and azimuth angles of satellites from ephemeris data.
    \item \emph{No nulling}: A baseline obtained by setting $\lambda = 0$ in $(\ref{eq:bf_eq})$, corresponding to SVD-based beamforming, which does not perform any interference nulling. 
\end{itemize}
    %
    %
\subsubsection*{Satellite uplink INR}

\begin{figure}[!t]
\centering
\subfloat[][CDF of the INR on the satellite uplink.]
{\includegraphics[width =0.95\columnwidth]{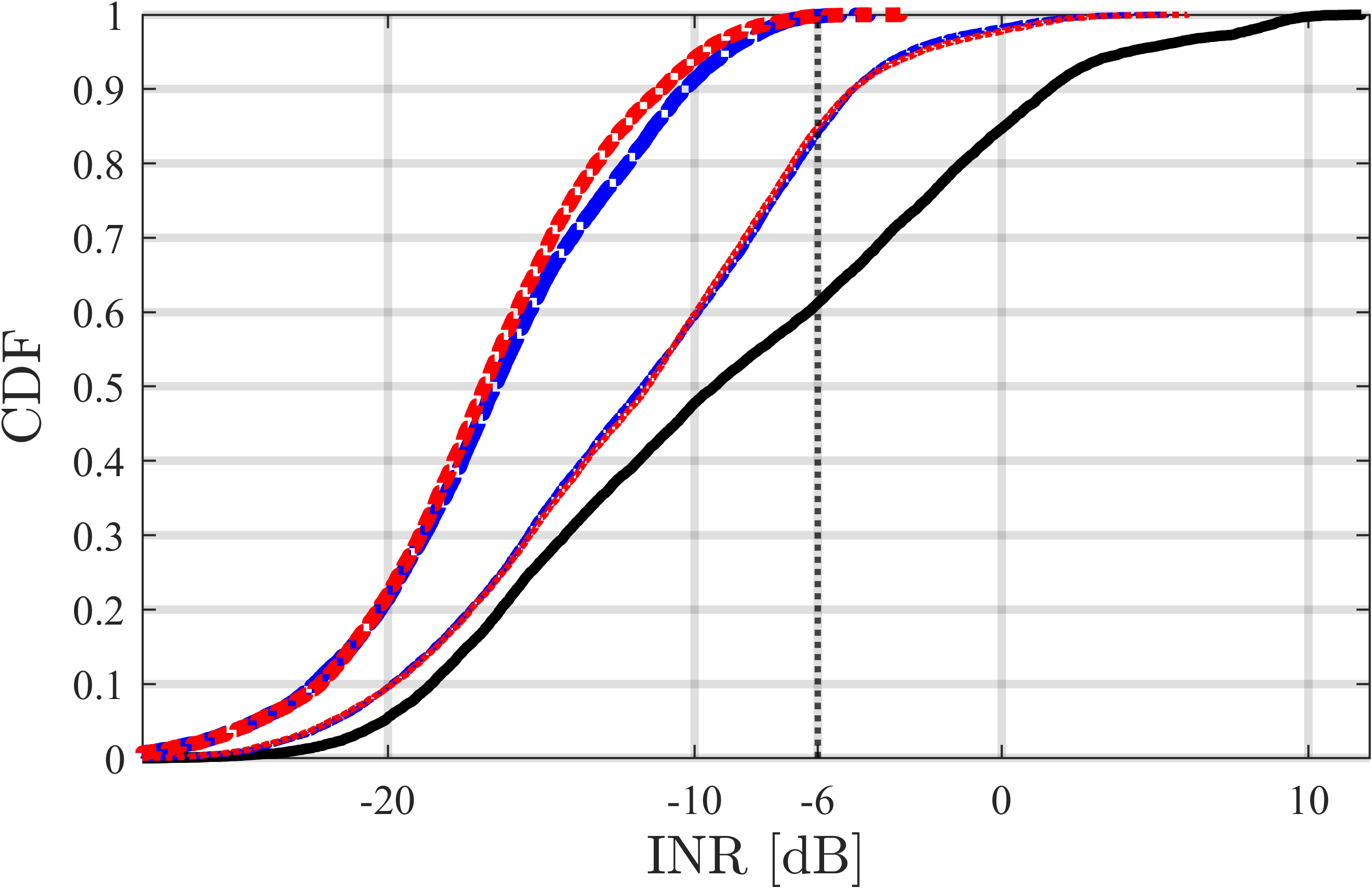} \label{fig:inr_cdf}}\\
\subfloat[][CDF of the SNR loss on the terrestrial downlink.]
{\includegraphics[width =0.95\columnwidth]{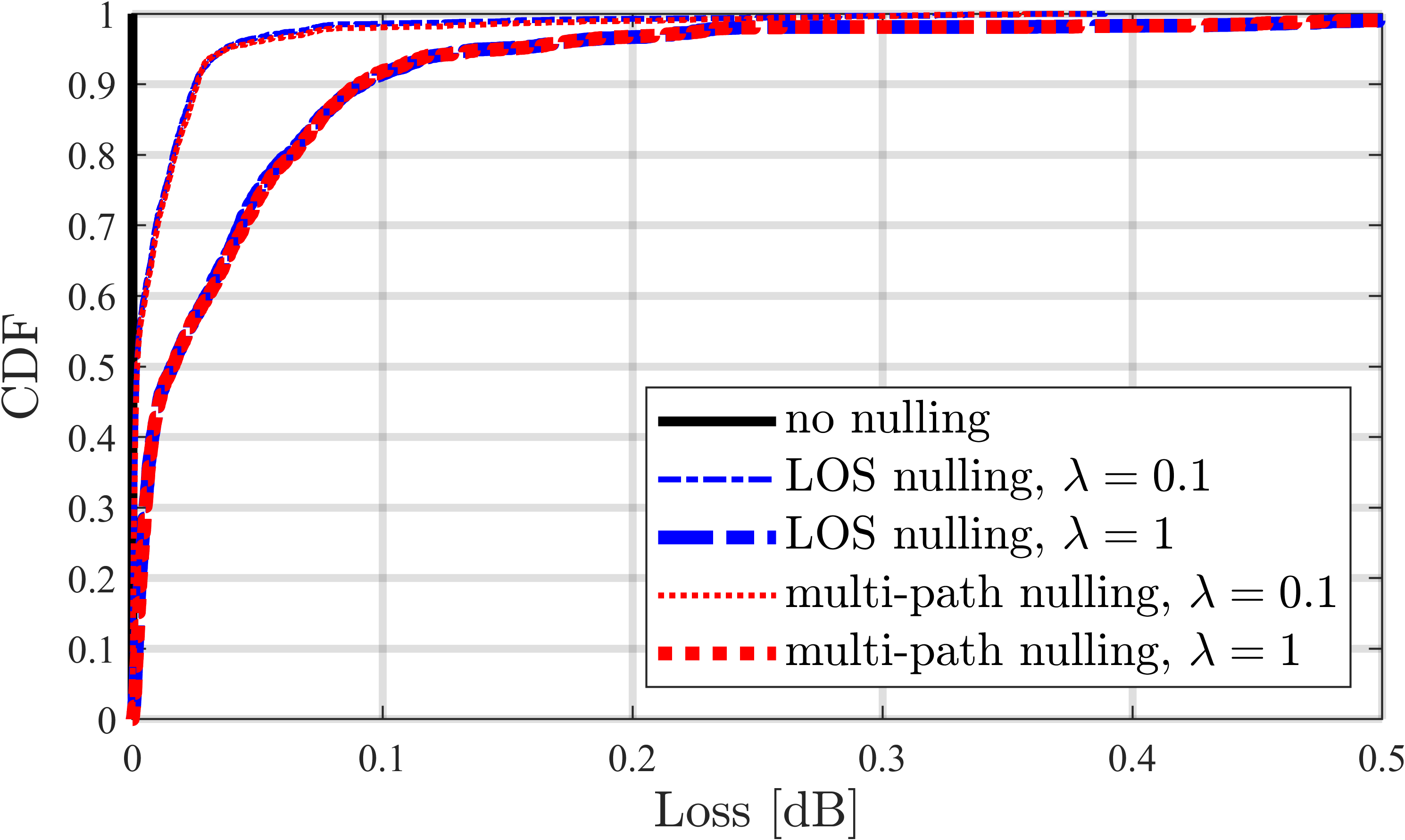} \label{fig:gloss_cdf}}
\caption{CDF of (a) the INR on the satellite uplink, and (b) the corresponding SNR loss on the terrestrial downlink.}
\label{fig:inr_loss_nbs10}
\end{figure}

Fig.~\ref{fig:inr_cdf} shows the CDF of the INR on the satellite link, defined in (\ref{eq:inr}), for the above three beamforming schemes. When interference nulling is performed, two values of $\lambda$ are considered, namely $0.1$ and $1$. 
In all cases, interference nulling (blue and red curves) is highly beneficial to the satellite uplink, which otherwise incurs $\text{INR} >$ \SI{-6}{dB} nearly $40\%$ of the time (black curve). With a sufficiently large regularization parameter, i.e., $\lambda\geq1$, both nulling schemes consistently keep the INR below \SI{-6}{dB}, thus guaranteeing a satellite uplink SNR degradation $\rho^{\rm S} <$ \SI{1}{dB}. 
Importantly, the proposed LOS interference nulling exhibits nearly the same performance as full multi-path nulling, making the former highly preferable due to its greater implementation feasibility. Indeed, non-LOS interference paths in rural area, such as those caused by diffraction, seem to cause negligible interference to the satellite uplink and do not justify tracking the multi-path channel for nulling purposes.  

\subsubsection*{Terrestrial downlink SNR degradation}

Similarly, Fig.~\ref{fig:gloss_cdf} shows the CDF of the SNR loss $\rho^{\rm T}$ on the terrestrial dowlink for the three different nulling schemes and two values of the regularization parameter $\lambda$. 
Observing Fig.~\ref{fig:gloss_cdf} alongside Fig.~\ref{fig:inr_cdf} confirms the role played by $\lambda$, whereby higher values (thicker dashed blue/red curves) reduce the INR on the satellite uplink while increasing the SNR degradation on the terrestrial downlink. 
Fig.~\ref{fig:gloss_cdf} also shows that even for $\lambda = 1$ (which guarantees very low INR on the satellite uplink, as shown in Fig.~\ref{fig:inr_cdf}) the proposed LOS interference nulling incurs an SNR loss on the terrestrial downlink that remains below \SI{0.1}{dB} $90\%$ of time.
Altogether, the results in Fig.~\ref{fig:inr_loss_nbs10} suggest that, by choosing a sufficiently large value of $\lambda$, nulling the LOS interference paths can successfully eliminate terrestrial-to-satellite interference with a negligible terrestrial performance loss.

\section{Conclusion}

In this paper, we studied spectrum sharing between terrestrial network downlink and LEO satellite uplink in the upper mid-band. We addressed the critical challenge of interference from terrestrial BSs, particularly through LOS sidelobe transmissions, by introducing a novel interference mitigation technique. This technique optimizes terrestrial beamforming by creating spatial nulls onto certain channel directions, which can be computed using readily available satellite ephemeris data. Our system-level simulations, which include tracking Starlink LEO satellites over rural Colorado, confirmed that our method can effectively prevent degradation of the satellite uplink while maintaining terrestrial downlink performance. Possible extensions of our work include considering multi-user MIMO terrestrial transmissions in the upper mid-band and generalizing the proposed interference nulling technique accordingly.




\bibliographystyle{IEEEtran}
\bibliography{bibl.bib}
\end{document}